\documentclass[%
 reprint,
 amsmath,amssymb,
 aps,
]{revtex4-1}
\usepackage{amsmath,mathtools}
\usepackage{subfigure}
\usepackage{graphicx}% Include figure files
\usepackage{dcolumn}% Align table columns on decimal point
\usepackage{bm}% bold math
\usepackage{float}
\usepackage[section]{placeins}
\begin{document}

\title{Nucleon shadowing effects in  $Cu+Cu$ and $Au+Au$  collisions at RHIC within the HIJING code}% Force line breaks with \\
%\thanks{A footnote to the article title}%

\author{Khaled Abdel-Waged}
\altaffiliation{khelwagd@yahoo.com.}%Lines break automatically or can be forced with \\
\author{Nuha Felemban}
% \email{Second.Author@institution.edu}
\affiliation{%
 Physics Department, Faculty of Applied Science, Umm Al-Qura university,\\
 P.O. Box (10471), Makkah, 21955, Saudi Arabia
}%

\begin{abstract}
The centrality dependence of pseudorapidity density of charged particles ($ dN_{ch}/d\eta $)  in $Cu+Cu$ ($Au+Au$) collisions at RHIC energy of  $ \sqrt{s_{_{\rm NN}}}=22.4$ , $62.4 $ and $200$  ($19.6$, $62.4$ and $200$) GeV, is investigated within an improved HIJING code. The standard HIJING model is enhanced by a prescription for collective nucleon-nucleon ($NN$) interactions and more modern parton distribution functions. The collective $NN$-interactions are used to induce both  cascade and nucleon shadowing effects. We find collective cascade broadens the pseudorapidity distributions in the tails (at $ |\eta|> y_{\rm beam} $) above $25-30\%$ collision centrality to be consistent with the $ dN_{ch}/d\eta $ data at $ \sqrt{s_{_{\rm NN}}}=19.6$, $22.4$, $62.4 $ GeV. The overall contribution of nucleon shadowing is shown to depress the whole shape of $ dN_{ch}/d\eta$ in the primary interaction region (at $ |\eta|< y_{\rm beam} $) for semiperipheral ($ 20-25 $ \%) and peripheral ($\ge 35-40$\%) $Cu+Cu$ ($Au+Au$) interactions at $ \sqrt{s_{_{\rm NN}}}=200$ GeV, in accordance with the PHOBOS data.   
\end{abstract}

%\pacs{Valid PACS appear here}% PACS, the Physics and Astronomy
                             % Classification Scheme.
%\keywords{Suggested keywords}%Use showkeys class option if keyword
                              %display desired
\maketitle

%\tableofcontents

\section{\label{sec:level1}INTRODUCTION}

The primary nucleon-nucleon ($NN$) interaction is the basic component for all microscopic transport models of light/heavy ion collisions \cite{1,2,3,4,5,6,7,8}. Due to the stopping mechanisms of initial $NN$-interactions, quark/gluon strings (soft scatterings) and new state of matter comprising of strongly interacting quarks and gluons (hard scatterings) could emerge at Relativistic Heavy Ion Collider (RHIC) energies. It has been observed that \cite{dis1, dis2} at the partonic level, the parton distribution functions (PDFs) in a primary interacting nucleon are subject to significant nuclear effects (so-called parton shadowing). A number of phenomenological approaches are thus developed to address the nuclear modification of PDFs which span a wide region of Bjorken $x$ \cite{1, pdf1,pdf2,pdf3,pdf4}.

HIJING \cite{1} is a microscopic hadronic model which treats $NN$-collision as a two component geometrical model of hard (with minijet production) and soft interactions. The hard component is characterized by a transverse momentum ($ p_{\scriptsize_T} $) larger than a cutoff scale ($ p_{0} $) and is  evaluated by perturbative QCD (pQCD) using the parton distribution function (PDF) in a nucleon. The soft interactions $ (p_{\scriptsize_T}< p_{0}) $ (non-pQCD) are modeled by the formation and fragmentation of quark-gluon strings. For proton-nucleus ($p+A$) and nucleus-nucleus ($A+B$) collisions, HIJING implements an eikonal formalism to determine the probability of collision, elastic or inelastic, and the number of jets produced in each binary $NN$-collision. In addition, HIJING takes into account both parton shadowing and jet quenching, which are shown to influence the flux of partons and, in turn, the description of initial particle production.

Although HIJING is mainly designed to explore nuclear medium effects that may occur in RHIC and LHC energy light/heavy ion collisions \cite{1,2}, it does not take into account the interactions of the primary nucleons with the dense nuclear medium (in short, nucleon shadowing), and the energy loss they suffer before applying shadowing on the partonic structure of the nucleon. The missing systematic inclusion of nucleon shadowing and the associated cascade effects should affect the relative contribution of initial particle production with centrality in light/heavy ion collisions.

Recently \cite{kh1,kh2}, we have presented a model for $p+A$ and $A+B$ collisions at LHC energies based on the available HIJING1.383 model with an updated modern sets of PDFs \cite{martin}. The main improvement in this model is implementing the collective cascade in a nucleus to get the nuclear modifications of nucleons (nucleon shadowing). This improved HIJING model (in short, ImHIJING/Cas) successfully reproduces the measured pseudorapidity density of charged particles ($ dN_{ch}/d\eta $) in $p+Pb$ collisions at $\sqrt{s_{_{\rm NN}}}=5.02$ TeV.  The largest nucleon shadowing effect is found to be contained in the pseudorapidity region of $3< \eta_{lab} <4$ \cite{kh1}. As for $Pb+Pb$ collisions at $\sqrt{s_{_{\rm NN}}}=2.76$ TeV, the model is shown to depress the pseudorapidity density at central rapidity, independent of the collision centrality, while enhancing the charged particles at large pseudorapidity ($|\eta| >8$) \cite{kh2}.  

However, in the previous analysis \cite{kh1,kh2} a single parameter set was used, in the collective cascade model, to reproduce the measured  $ dN_{ch}/d\eta $ at mid-pseudorapidity region for both $p+Pb$ and $Pb+Pb$ collisions at LHC energy, and no constraints of the model parameters from pure soft $NN$ processes were given.  

In this paper and in light of PHOBOS data at RHIC \cite{9}, we explore different parameters of the collective cascade model that are important for a proper description of the whole shape of $ dN_{ch}/d\eta $ at different centrality bins. After having constrained the model parameters from $Au+Au$ collisions at the lowest RHIC energy ($ \sqrt{s_{_{\rm NN}}}=19.6$ GeV) we shall study nucleon shadowing and the associated cascade effects in $Au+Au$($Cu+Cu$) collisions at higher RHIC energies.  Unlike our previous studies \cite{kh1,kh2}, the nucleon shadowing, constrained by the lowest RHIC data, is found to have an impact-parameter dependence. This will have important implications on the study of other phenomena such as parton shadowing and parton energy loss  in heavy ion collisions, since they all depend on $NN$ interactions in nuclear environment.

The manuscript is organized as follows: in Section II, we describe the ImHIJING/Cas model, that implements nucleon shadowing and the associated cascade effects. Then, in Sec. III we use the model to analyze the PHOBOS \cite{9} results on $ dN_{ch}/d\eta $ as a function of collision centrality for  $Cu+Cu$ ($Au+Au$) collisions at $ \sqrt{s_{_{\rm NN}}}=22.4 $, $62.4$ and $200$ ($19.6$, $62.4$, $130$ and $200$) GeV. Finally, in Sec. IV we present our conclusions.

\section{\label{sec:level2}Description of the model}

Our calculations are based on the ImHIJING/Cas model presented in Refs. \cite{kh1, kh2}. Here we will concentrate on those points which are important for understanding the results discussed in Sec. III. 

\begin{figure*}[] 
 	\begin{center}
 		\includegraphics[width=\linewidth]{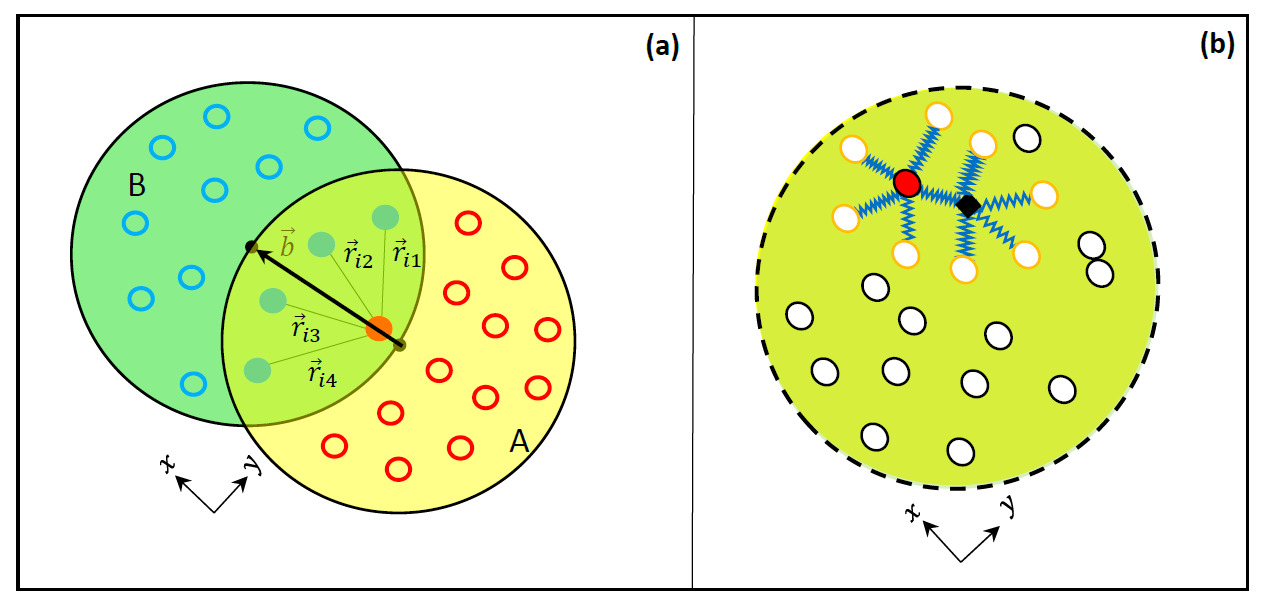}
 		\caption{(COLOR ONLINE)	
 			A schematic view of (a) primary and (b) secondary interactions in nucleus-nucleus collisions on the impact-parameter plane. All nucleons are shown as open circles and primary interacting nucleons are marked by closed circles. The wavy lines and the square point in (b) represent the set of individual Reggeon exchanges and the Reggeon interaction vertex, respectively. 
 		}
 		\label{f0}
 	\end{center}
 \end{figure*}

The ImHIJING/Cas is a hybrid model which combines the available HIJING 1.383 code \cite{1} with more modern parton distribution functions \cite{martin} and a collective cascade recipe \cite{kh3, kh4}. In ImHIJING/Cas, nucleons taking part in the primary interactions (see Fig.\ref{f0}) suffer both parton and nucleon shadowing effects.ImHIJING/Cas determines nucleon shadowing of the primary interacting nucleons of the projectile ($A$) and target ($B$) nuclei  at a given impact parameter, while both the radial and  mass number dependent parton shadowing are adopted from standard HIJING \cite{1}. The collective cascade recipe is used to calculate both nucleon shadowing and cascading in $A+B$ collisions.

As described by the Reggeon theory inspired model \cite{kh3, kh4}, nucleon shadowing arises due to simultaneous interactions of a primary nucleon with other non-interacting nucleons from the target/projectile, see Fig.\ref{f0}b. In $p+A$ and $A+B$ collisions, the probability to involve the $i^{th}$ non-interacting (secondary) nucleon by the $j^{th}$ primary interacting nucleon, at a distance $ r_{ij} $, is taken as 
\begin{equation}\label{eq1} 
	w=C \exp(-r_{ij}^{2}/r_{0}^{2}),
\end{equation}
where $r_{0}=1.2$ fm  is the mean interaction radius and  $r_{ij}=\sqrt{(b_{x}+x_{i}-x_{j} )^{2}+(b_{y}+y_{i}-y_{j} )^{2} }$  is the transverse distance of the interacting nucleon pair ($i$ and $j$),$ b_{x(y)}$ and $x_{i(j)}, y_{i(j)} $are the components of the impact parameter vector and the coordinates of the pair measured from their own nucleus. The parameter $C$ determines the strength of nucleon shadowing and the associated cascade effects. If the number of newly involved nucleons is not zero, then such a nucleon can involve another secondary nucleon and so on, see Fig.\ref{f0}b. The eikonal formalism, as implemented in HIJING \cite{1}, is used to determine the primary interacting nucleons of the projectile ($A$) and target ($B$) nuclei  at a given impact parameter.

The energy/momentum is shared between the primary and accompanying secondary nucleons. To take this into account, we ascribe to each wounded (primary or secondary) nucleon a transverse momentum distribution according to the law 

\begin{equation}\label{eq2}
P(p^{\prime}_{\scriptsize_{Ti}})\propto \prod_{i=1}^{N_{A(B)}} \exp[- p^{\prime 2}_{\scriptsize_{Ti}}/<p_{\scriptsize_{\scriptsize_{T}}}^{2}>]  \delta(\sum_{i=1}^{N_{A(B)}} p^{\prime}_{\scriptsize_{Ti}}) d p^{\prime}_{\scriptsize_{Ti}}                                                   
\end{equation} 
where $ N_{A(B)}$ is the number of wounded nucleons from the projectile/target.

The fractional momentum of the wounded nucleon is chosen as  

\begin{equation}\label{eq3}
	P(x_i^{\prime +}) \varpropto \prod_{i=1}^{N_{A(B)}} 
	\exp[-\frac{(x_{i}^{\prime +}-\frac{1}{N_{A}})^2}{d^{2}}]\\
	\delta(1-\sum_{i=1}^{N_{A(B)}} x_{i}^{\prime +} )  d x_{i}^{\prime +},                                                                        
\end{equation}

with a mean $ \mu=1/N_{A(B)} $ and a width given by \cite{26,27}
\begin{equation}\label{eq4}  
	d=\gamma/N_{A(B)}                                                           
\end{equation}

The energy-momentum conservation between the primary and accompanying secondary nucleons is satisfied as follows.

	\begin{equation}\label{5}
		\begin{split}
			\sum_{i=1}^{N_{A}} E^{\prime}_{i} +\sum_{j=1}^{N_{B}} E^{\prime}_{j}&=
			\frac{W^{\prime +}_{A}}{2}+ \frac{1}{2 W^{\prime +}_{A}} \sum_{i=1}^{N_{A}}\frac{m^{\prime 2}_{\scriptsize_{Ti}}}{x^{\prime +}_{i}}\\ &+\frac{W^{\prime -}_{B}}{2}
			+\frac{1}{2 W^{\prime -}_{B}} \sum_{j=1}^{N_{B}}\frac{\mu^{\prime 2}_{\scriptsize_{Tj}}}{x^{\prime -}_{j}}\\
			&=E_{A}^{0}+E_{B}^{0},
		\end{split}
	\end{equation}
	\begin{equation}\label{6}
		\begin{split}
			\sum_{i=1}^{N_{A}} p^{\prime }_{zi} +\sum_{j=1}^{N_{B}} q^{\prime}_{zj}&=
			\dfrac{W^{\prime +}_{A}}{2}-\frac{1}{2 W^{\prime +}_{A}}\sum_{i=1}^{N_{A}}\frac{m^{\prime 2}_{\scriptsize_{Ti}}}{x^{\prime +}_{i}}\\
			&-\frac{W^{\prime -}_{B}}{2}
			+\frac{1}{2 W^{\prime -}_{B}}\sum_{j=1}^{N_{B}}\frac{ \mu^{\prime 2}_{\scriptsize_{Tj}}}{ x^{\prime -}_{j}}\\
			&=p_{zA}^{0}+q_{zB}^{0},
		\end{split}
	\end{equation}
	and
	\begin{equation}\label{7}
		\sum_{i=1}^{N_{A}} p^{\prime}_{\scriptsize_{Ti}}+\sum_{j=1}^{N_{B}} q^{\prime}_{\scriptsize_{Tj}}=0,
	\end{equation}
where
$W_A^+=\sum_{i=1}^{N_A}(E_{i}+p_{zi}),$ and $W_{B}^{-}=\sum_{j=1}^{N_{B}}(E_{j}-q_{zj})$. Here $E_{i} (E_{j}) $ and $ p_{zi} (q_{zj})$ are the initial energy and longitudinal momentum of  the $i^{th}$ ($j^{th}$) wounded nucleon. The corresponding total energy and momentum are given by $E^{0}_{A}=\sum_{i=1}^{N_{A}}E_i$ ($E^{0}_{B}=\sum_{j=1}^{N_{B}}E_j$), $ p_{z A}^{0}=\sum_{i=1}^{N_A}p_{zi}$ ($ q_{z B}^{0}=\sum_{j=1}^{N_B}q_{zj}$), respectively. The final momenta of the $ i^{th}$ and $j^{th}$ wounded nucleons are given by     
	
	\begin{equation}\label{eq8}
		p^{\prime}_{zi}=(W^{\prime +}_{A} x^{\prime +}_{i}-\frac{ m^{\prime 2}_{\scriptsize_{Ti}}} {x^{\prime +}_{i} W^{\prime +}_{A}})/2,
	\end{equation}
	
	\begin{equation}\label{eq9}
		q^{\prime}_{zj}=-(W^{\prime -}_{B} x^{\prime -}_{j}- \frac{ \mu^{\prime 2}_{\scriptsize_{Tj}}}{ x^{\prime -}_{j} W^{\prime -}_{B}})/2,
	\end{equation}
	where $m^{\prime 2}_{\scriptsize_{Ti}}={m}_{i}^{2}+ p^{\prime 2}_{\scriptsize_{Ti}}$, $\mu^{\prime 2}_{\scriptsize_{Tj}}=\mu_{j}^{2}+{q}_{\scriptsize_{Tj}}^{2}$, and $m_i (\mu_{j})$ is the mass of the $i^{th} (j^{th})$ secondary interacting nucleon from $A$($B$).

Thus, nucleons taking part in the primary interactions suffer energy loss due to cascading with other noninteracting ones, and the remaining energy is used to produce jets or excited strings according to the HIJING model. It should be noted that, the energy-momentum conservation of primary interacting nucleons is satisfied in HIJING. The energy and momentum carried away by jets in hard collisions is subtracted from that of nucleons, only the leftover energy-momentum of the nucleon is available for soft interactions. The energy-momentum exchange between nucleons in the soft collision also satisfies conservation laws.

 In order to calculate the effective jet cross section ($\sigma^{\rm{eff}}_{\rm{jet}}$) in $A+B$ collisions, we have to calculate the parton shadowing ($\alpha_{A} (r_{i})$) at the radial position of the nucleon ($r_{i}$) measured from its own nucleus center, with $r_{i}=\sqrt{x_{i}^{2}+y_{i}^{2}}$. In this work, the HIJING 2.0 parameterization of  parton shadowing is adopted \cite{2} 

\begin{equation}\label{eq10}
\alpha_{A} (r_{i} )=s_{q(g)}  (A^{1/3}-1) \: \dfrac{5}{3} \:   \: (1-\dfrac{r_{i}^{2}}{R_{A}^{2}})
\end{equation}
where $R_{A}=1.2 A^{1/3}$ is the nuclear radius. Here $s_{q(g)} $  is a single unknown parameter of the model that should be fixed from comparison to the measured data of the centrality dependence of charged particle pseudorapidity density in $A+B$ collisions. 

As a result of parton shadowing, the effective jet cross section, e.g., between  two nucleons $i$ and $j$ in $A+B$ collisions, $\sigma^{\rm{eff}}_{\rm{jet}}$,  is calculated by multiplying the jet cross section ($\sigma^{\rm{AB}}_{\rm{jet}}$) by  $\alpha_{A}(r_{i})R_{a/A}^{s}(x_{a}) f_{a/i} (x_{a}, p^{2}_{T} )\times \alpha_{B}(r_{j})R_{b/B}^{s}(x_{b}) f_{b/j} (x_{b}, p^{2}_{T}) $, where $f_{a(b)/i(j)} (x_{a(b)}, p^{2}_{T} )$ and $R^{s}_{a(b)/A(B)}$ are the parton distribution function and the mass number dependent shadowing factor. Thus the hard hadron production, which is associated with $\sigma^{\rm{eff}}_{\rm{jet}}$, becomes  affected by the radial distance, $r_{i(j)}$, of the nucleon with respect to its own nucleus.

In previous studies of $p+Pb$ and $Pb+Pb$ collisions at LHC energy  \cite{kh1, kh2}, a single set of parameters ($<p_{\scriptsize_{T}}^{2}>=0.5$ (GeV/c)$^{2}$, $ \gamma= 0.5$, $C=1$) was used in the collective cascade model to reproduce the measured $ dN_{ch}/d\eta$ at mid-pseudorapidity. In the present study of heavy ion collisions at RHIC energies, we vary the values of these parameters to fit the measured data. As shown below, a reasonable description of the measured  $ dN_{ch}/d\eta$ in the high-$\eta$ tail regions (at $ |\eta|> y_{\rm beam} $ ) is achieved if different collective cascade strengths   ($C=0.25-0.5$) are used. Also, an explicit impact parameter dependence of the width 

\begin{equation}\label{eq11}
	d=\gamma(c)/N_{A(B)}
\end{equation}
is assumed to take into account the impact parameter dependence of nucleon shadowing. In Eq.(\ref{eq11}) the centrality $c$ is related to the impact parameter by the empirical formula $c=\pi b^2/\sigma_{in}$ \cite{28} where $\sigma_{in}$ is the $A+B$ inelastic cross section.
 
Experimental measurements at different collision centralities could serve to disentangle the new collective cascade parameter $ \gamma(c) $. The parameter $ \gamma(c) $ which fits the measured 
centrality dependence of  $ dN_{ch}/d\eta $ at mid-pseudorapidity ($ |\eta|<1 $) in $Au+Au$  collisions at the lowest collision energies of $ \sqrt{s_{_{\rm NN}}} =19.6 $ GeV, takes the form
\begin{equation}\label{eq12}
 \gamma(c)=0.1+2c(1-\frac{3}{4}\sqrt{c}).
\end{equation}
      
\begin{figure*}[] 
 	\begin{center}
 		\includegraphics[width=\linewidth]{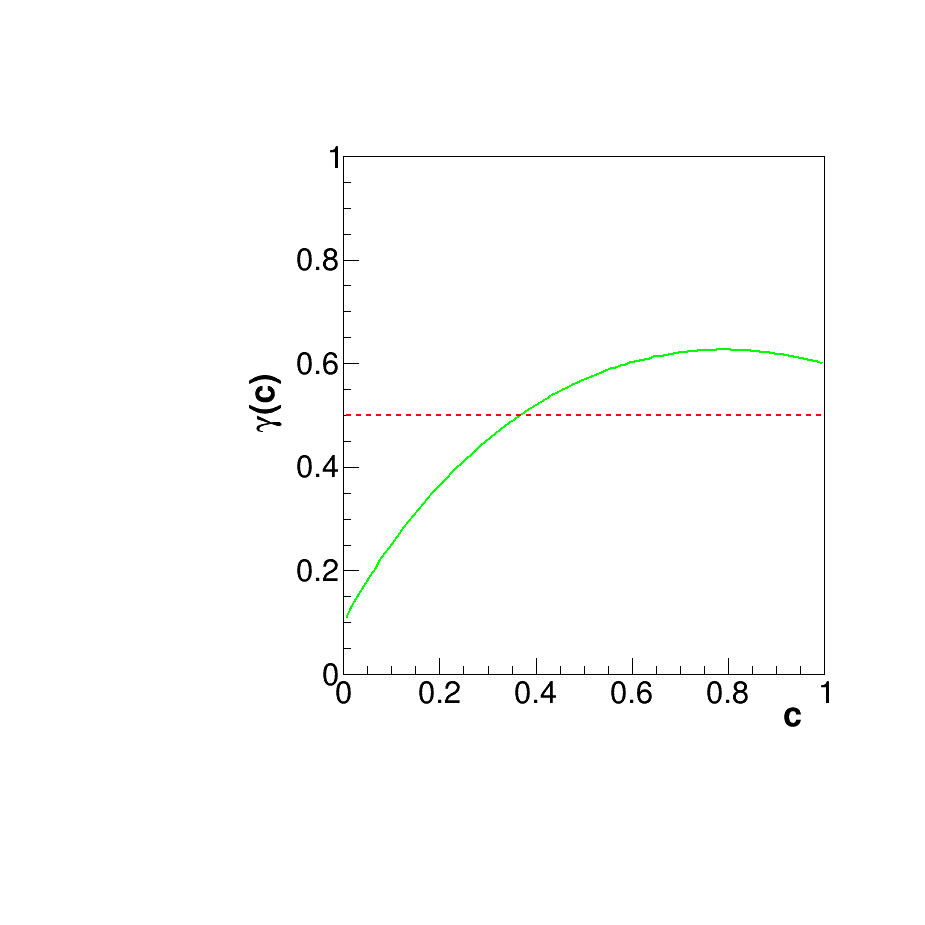}
 		\caption{(COLOR ONLINE)	
 			The parameter which controls the longitudinal fractional momentum of secondary interacting nucleons in the collective cascade model. The solid and short-dashed lines denote the new and old parameterizations as a function of collision centrality ($c$), respectively. 	
 		}
 		\label{f1}
 	\end{center}
 \end{figure*}

\begin{figure*}[] 
  	\begin{center}
  		\includegraphics[width=\linewidth]{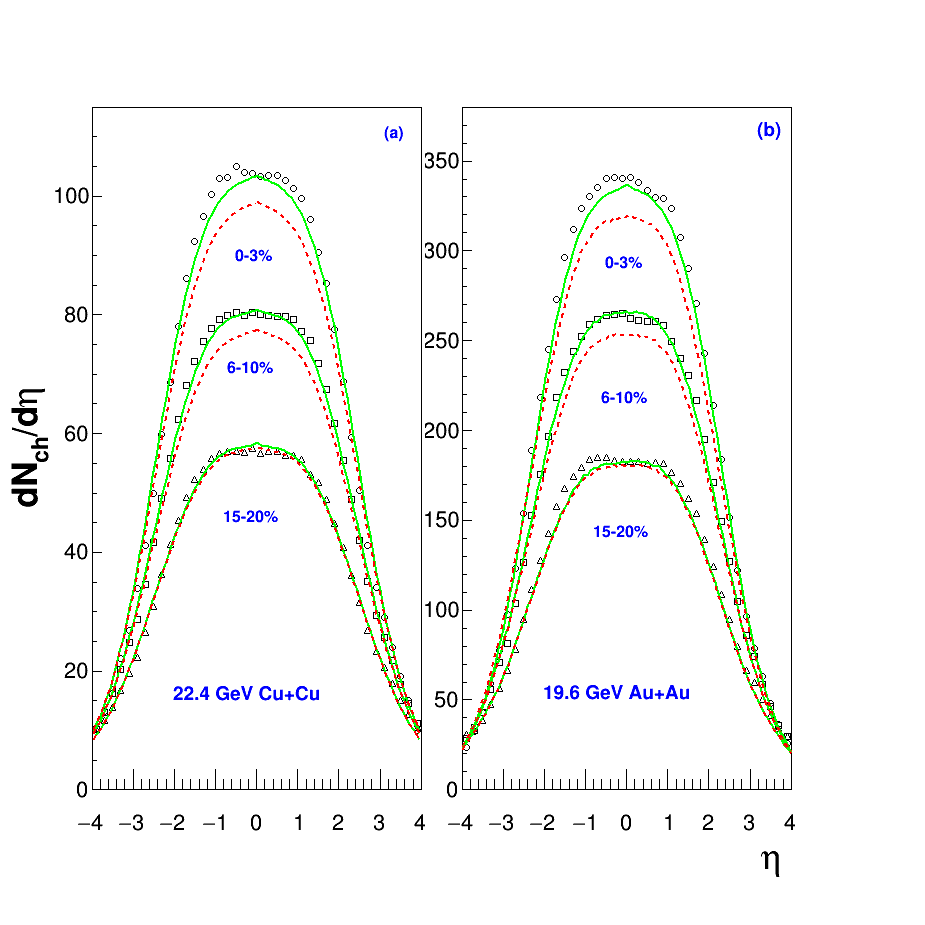}
  		\caption{(COLOR ONLINE)	
  		Pseudorapidity density distributions of charged particles as a function of  selected collision centralities: (a) for $Cu+Cu$ collisions at $ \sqrt{s_{_{\rm NN}}}=22.4 $ GeV,  and (b) for $Au+Au$ collisions at $ \sqrt{s_{_{\rm NN}}}=19.6 $ GeV. The open points denote PHOBOS data \cite{9}. The solid and short-dashed lines denote the ImHIJING/Cas calculations with $ \gamma(c) $ and $ \gamma=0.5 $, respectively (see text for details). 		
  		}
  		\label{f2}
  	\end{center}
  \end{figure*}

According to Eq.(\ref{eq12}) the width increases rapidly as the centrality increases and approaches nearly a plateau at $c>0.55$, with $\gamma\approx 0.1-0.2 $ and $0.6$ for central ($ c=0-10\% $) and peripheral ($ >50-60\%$) collisions, respectively; see Fig.\ref{f1}.
As one can see in Fig.\ref{f2}, the data could not be described by assuming only a constant width of $\gamma=0.5$ (dashed lines). On the other hand, the introduction of  $\gamma(c)$ (solid lines) in ImHIJING/Cas increases the level of $ dN_{ch}/d\eta $ in central ($0-10\%$) interactions, independent of the colliding system, and results in a better agreement with the data. This implies that  $ x_{i(j)}^{\prime \pm}(c) $ leads to larger fractional momentum of the $i^{th} (j^{th})$ wounded nucleon in central than peripheral collisions, as it should be.
 
Constrained by the model calculations of $ dN_{ch}/d\eta$ per centrality class for $Au+Au$ collisions at the lowest RHIC energy ($ \sqrt{s_{_{\rm NN}}}=19.6$ GeV) we will then be able to discuss possible nucleon shadowing effects in $A+B$ collisions at different RHIC energies.
 
In the numerical calculations, the ImHIJING/Cas is running in two modes, the cascade mode with different strengths $C$ of nucleon shadowing and the one that does not include nucleon shadowing ($C=0$). In both modes, the shadowing on PDFs are implemented, Eq.(\ref{eq10}), for the studied reactions at $\sqrt{s_{_{\rm NN}}} \geq 62.4$ GeV. Thus, the differences observed in the final results of  ImHIJING/Cas are regarded as evidence of nucleon shadowing effects at RHIC. Note that, in all calculations, the default HIJING1.383 parameters are selected and no adjustments are attempted.

\section{\label{sec:level3}RESULTS AND DISCUSSION}
In this section we present the predictions of the  ImHIJING/Cas code along with the recent measurements of  PHOBOS (for  $0-55\%$ event centralities) results on pseudorapidity distributions of charged particles emitted in heavy ion collisions over a wide energy range. The PHOBOS data cover $Cu+Cu$ collisions at $NN$ center of mass energy, $ \sqrt{s_{_{\rm NN}}} $, of $22.4$, $62.4$ and $200$ GeV, $Au+Au$ at $19.6$, $62.4$ and $200$ GeV \cite{9}.

\begin{figure*}[] 
    	\begin{center}
    		\includegraphics[width=\linewidth]{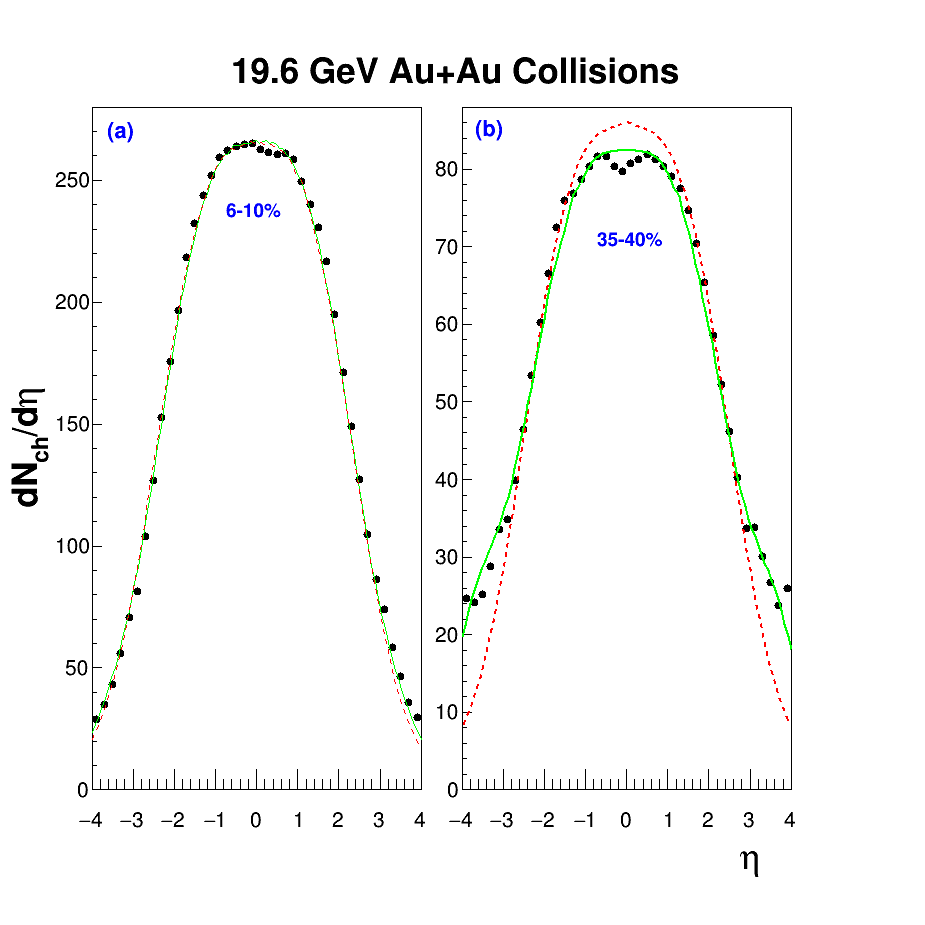}
    		\caption{(COLOR ONLINE)	
   	Pseudorapidity density distributions of charged particles in $Au+Au$ collisions at $ \sqrt{s_{_{\rm NN}}}=19.6 $ GeV: (a) for $6-10$\% and (b)  $35-40$\% centrality intervals. The filled points denote PHOBOS data \cite{9}. The solid and short-dashed lines denote the ImHIJING/Cas calculations using $\gamma(c)$ with $C=0.5$ and $C=0$, respectively. 
    		}
    		\label{f4}
    	\end{center}
    \end{figure*}

\begin{figure*}[] 
     	\begin{center}
     		\includegraphics[width=\linewidth]{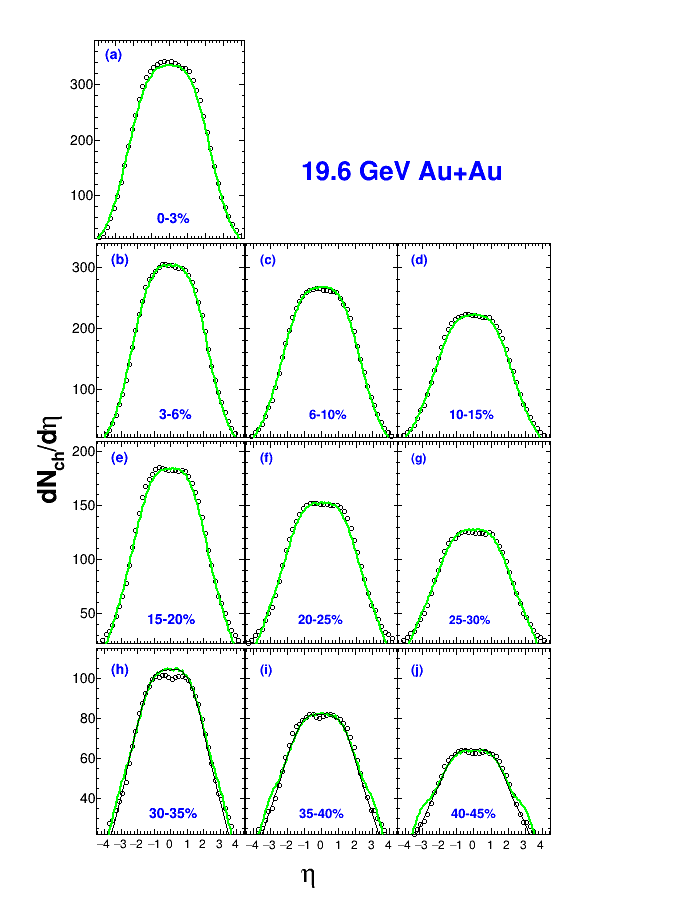}
     		\caption{(COLOR ONLINE)	
     Pseudorapidity density distributions of charged particles in $Au+Au$ collisions at $ \sqrt{s_{_{\rm NN}}}=19.6 $ GeV for ten centrality bins. The experimental data (open points) are from the PHOBOS experiment \cite{9}. The thick and thin solid lines denote the ImHIJING/Cas calculations  with cascade strengths of $C=0.5$ and $0.25$, respectively.		
     		}
     		\label{f5}
     	\end{center}
     \end{figure*}

\begin{figure*}[] 
      	\begin{center}
      		\includegraphics[width=\linewidth]{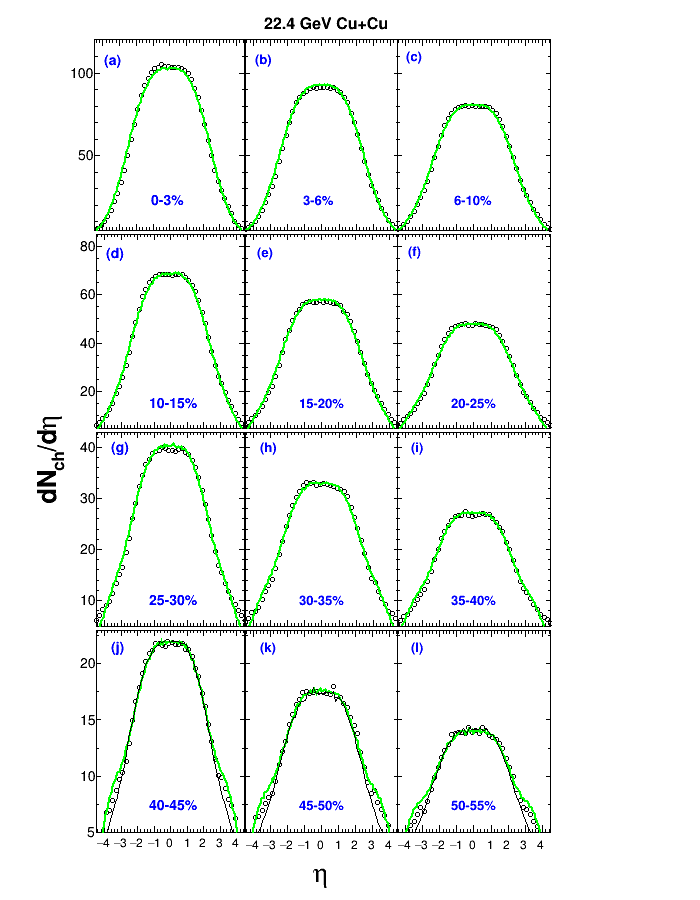}
      		\caption{(COLOR ONLINE)	
      		Same as Fig.\ref{f5}, but for $Cu+Cu$ collisions at $ \sqrt{s_{_{\rm NN}}}=22.4 $ GeV and  twelve centrality bins.	
      		}
      		\label{f6}
      	\end{center}
      \end{figure*}

\begin{figure*}[] 
       	\begin{center}
       		\includegraphics[width=\linewidth]{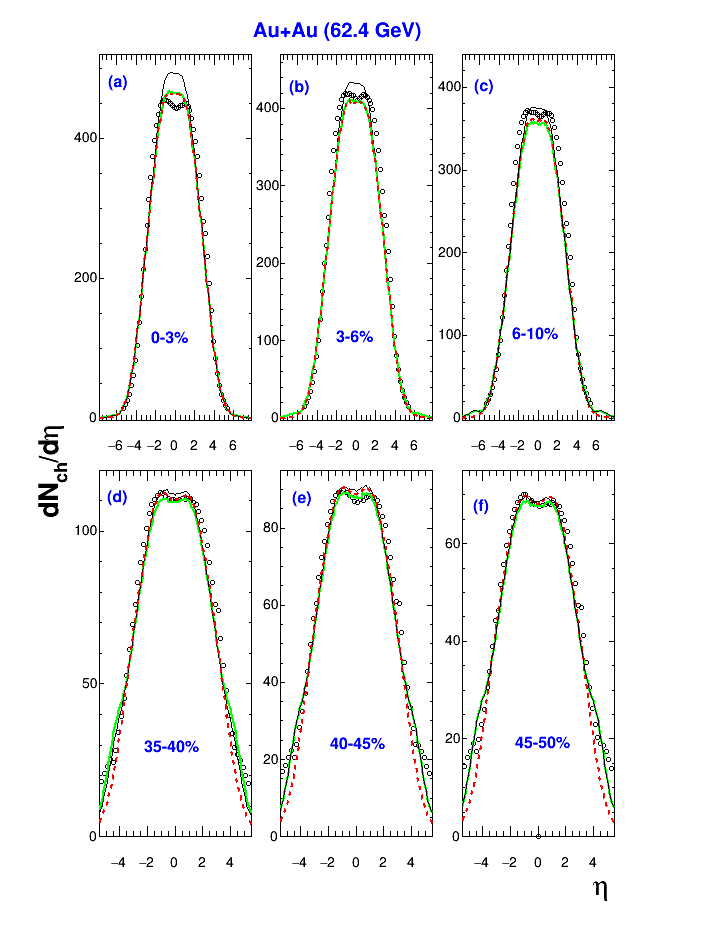}
       		\caption{(COLOR ONLINE)	
       		Pseudorapidity density distributions of charged particles in central (top panels) and peripheral (bottom panels) $Au+Au$ collisions at  $ \sqrt{s_{_{\rm NN}}}= 62.4$ GeV. The  experimental data (open points)  are from the PHOBOS experiments \cite{9}. The solid and short-dashed lines denote the ImHIJING/Cas calculations with $C=0.25$ and $0$, respectively, with $ s_{q(g)}=0.014$ in both cases. The thin solid lines are the ImHIJING/Cas calculations with $C=0.25$ and $ s_{q(g)}=0.0$.   	
       		}
       		\label{f7}
       	\end{center}
       \end{figure*}

\begin{figure*}[] 
        	\begin{center}
        		\includegraphics[width=\linewidth]{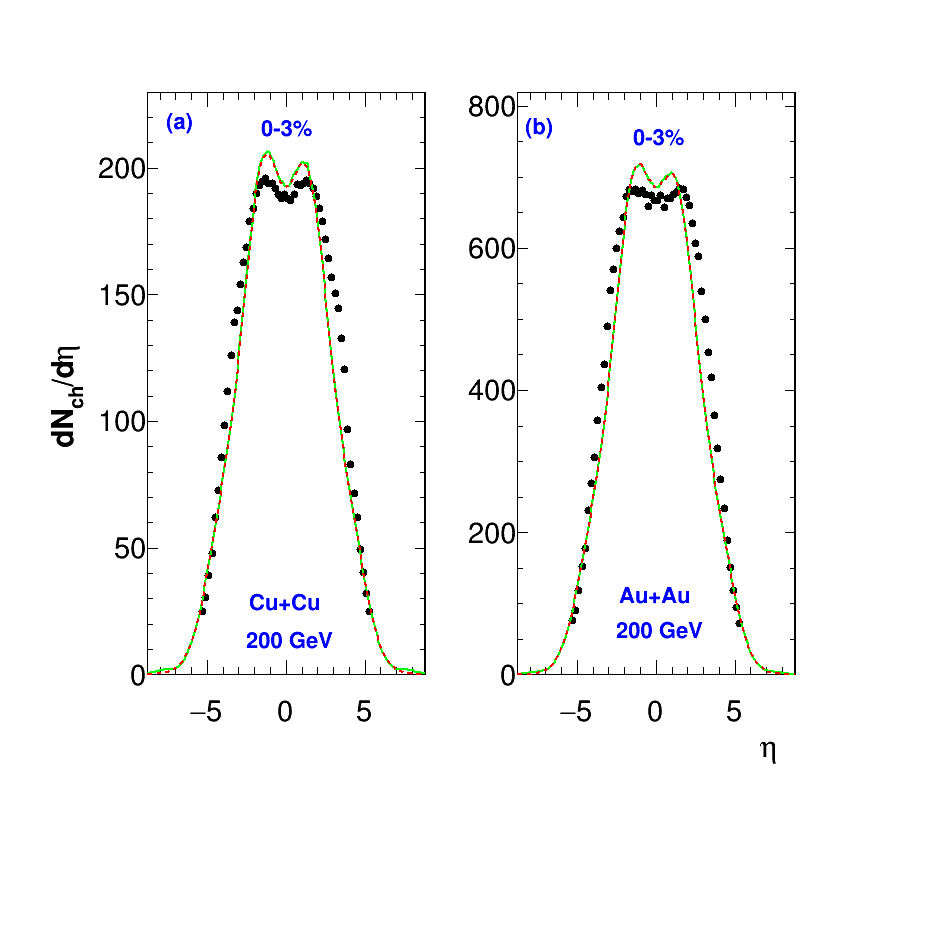}
        		\caption{(COLOR ONLINE)	
Pseudorapidity density distributions of charged particles in the most central ($0-3\%$); (a) $Cu+Cu$ and (b) $Au+Au$ collisions at  $ \sqrt{s_{_{\rm NN}}}= 200$ GeV. The  experimental data (closed points)  are from the PHOBOS experiments \cite{9}. The solid and short-dashed lines denote the ImHIJING/Cas calculations with cascade strengths of $ C=0.5 $ and $0$, respectively.		
        		}
        		\label{f8}
        	\end{center}
        \end{figure*}

\begin{figure*}[] 
         	\begin{center}
         		\includegraphics[width=\linewidth]{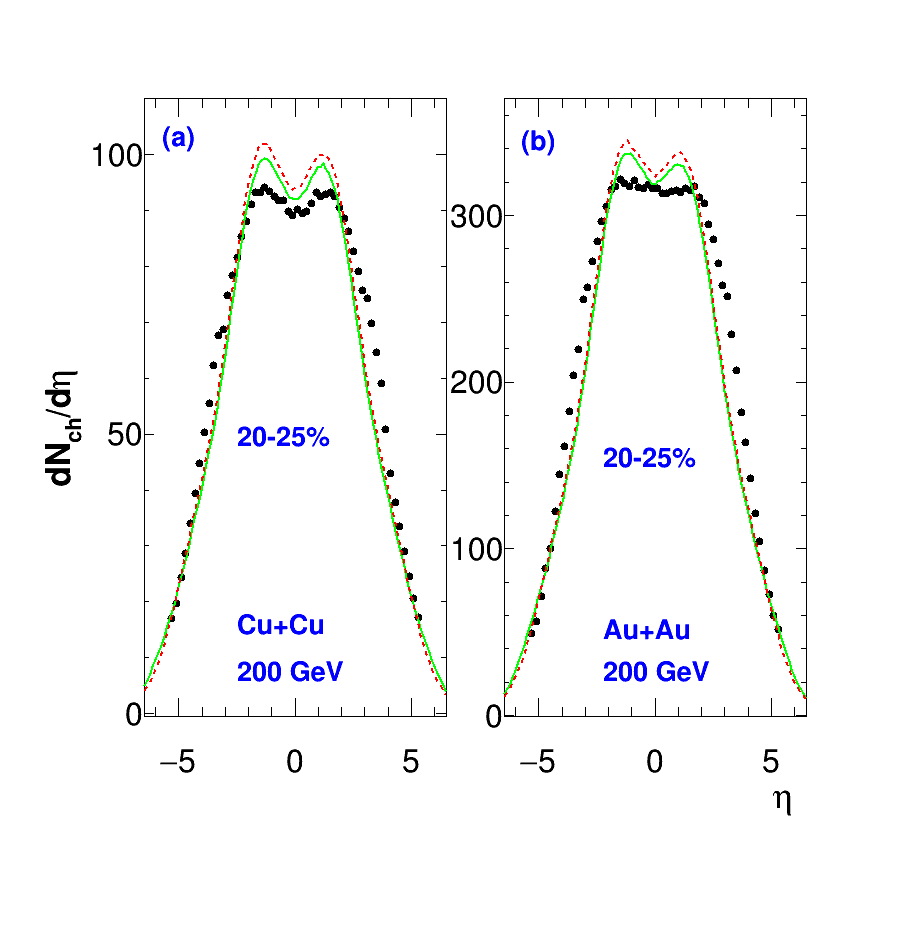}
         		\caption{(COLOR ONLINE)	
         Same as Fig.\ref{f8}, but for semiperipheral ($20-25\%$)  $Cu+Cu$ and $Au+Au$ collisions at $ \sqrt{s_{_{\rm NN}}}=200$ GeV.
         		}
         		\label{f9}
         	\end{center}
         \end{figure*}

\begin{figure*}[] 
          	\begin{center}
          		\includegraphics[width=\linewidth]{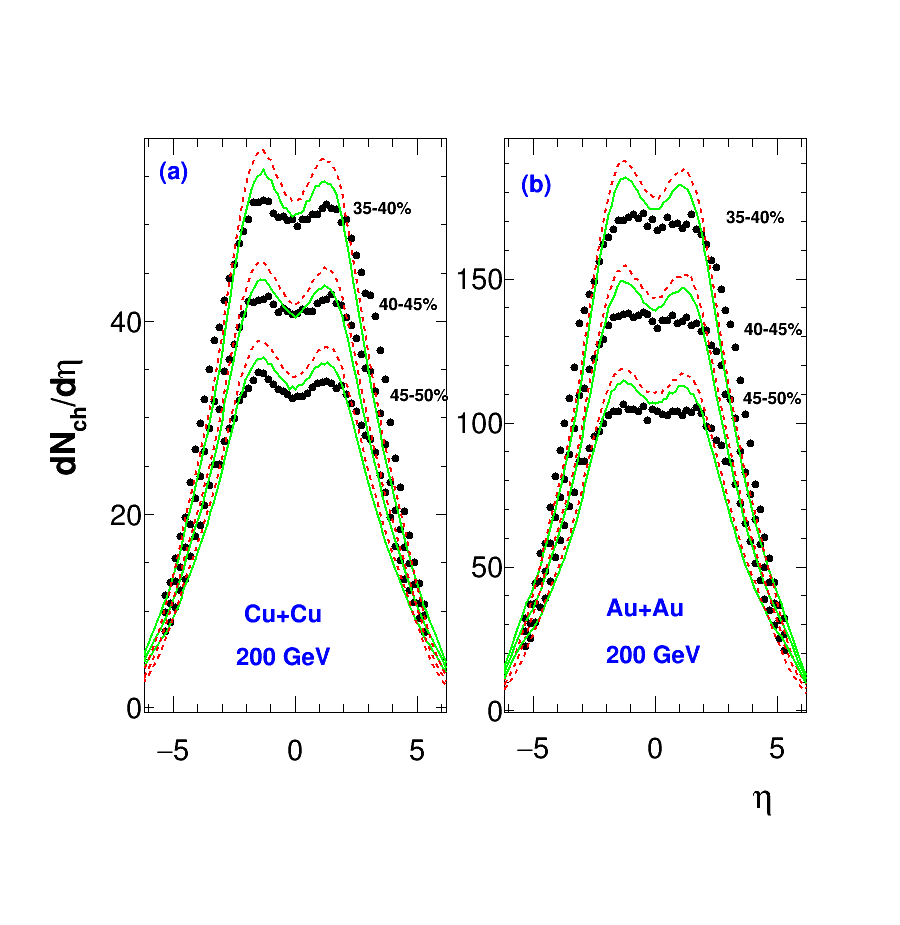}
          		\caption{(COLOR ONLINE)	
       Same as Fig.\ref{f8}, but for peripheral ($\ge 35-40\%$) $Cu+Cu$ and $Au+Au$ collisions at $ \sqrt{s_{_{\rm NN}}}= 200$ GeV. 			
          		}
          		\label{f10}
          	\end{center}
          \end{figure*}

The measured $ dN_{ch}/d\eta$ distributions (see Figs. \ref{f4}-\ref{f10}) are composed of three components. At mid-pseudorapidity region, a change of the pseudorapidity density distribution appears, depending on both the centrality and collision energy. The change is from a Gaussian to a double Gaussian shape, due to Jacobian. In addition, two components exist for all of the spectra: one is a smooth fall-off  below beam rapidity, $ y_{\rm beam} $, the other is a large high-$\eta$ tail  above $ y_{\rm beam} $. The smooth fall-off pseudorapidity region increases in range with collision energy. The charged particle production in both the mid- and smooth fall-off  pseudorapidity regions is strongly dominated by primary interacting nucleons. Whereas the tails, developed as one moves towards peripheral collisions and lower collision energies, are attributable to charged particles emitted from the spectator regions of projectile/target. Below, we investigate nucleon shadowing and the associated cascade effects on the three components of $ dN_{ch}/d\eta $ by employing ImHIJING/Cas. Because $ dN_{ch}/d\eta $ are measured in minimum bias, we generate $ 10^{4}$events for each centrality bin. 

We use for the different centrality classes the range of impact parameter based on PHOBOS results. The derived $b$-range for all studied reactions are listed in Tables I and II of the Appendix. 
 
\begin{figure*}[] 
   	\begin{center}
   		\includegraphics[width=\linewidth]{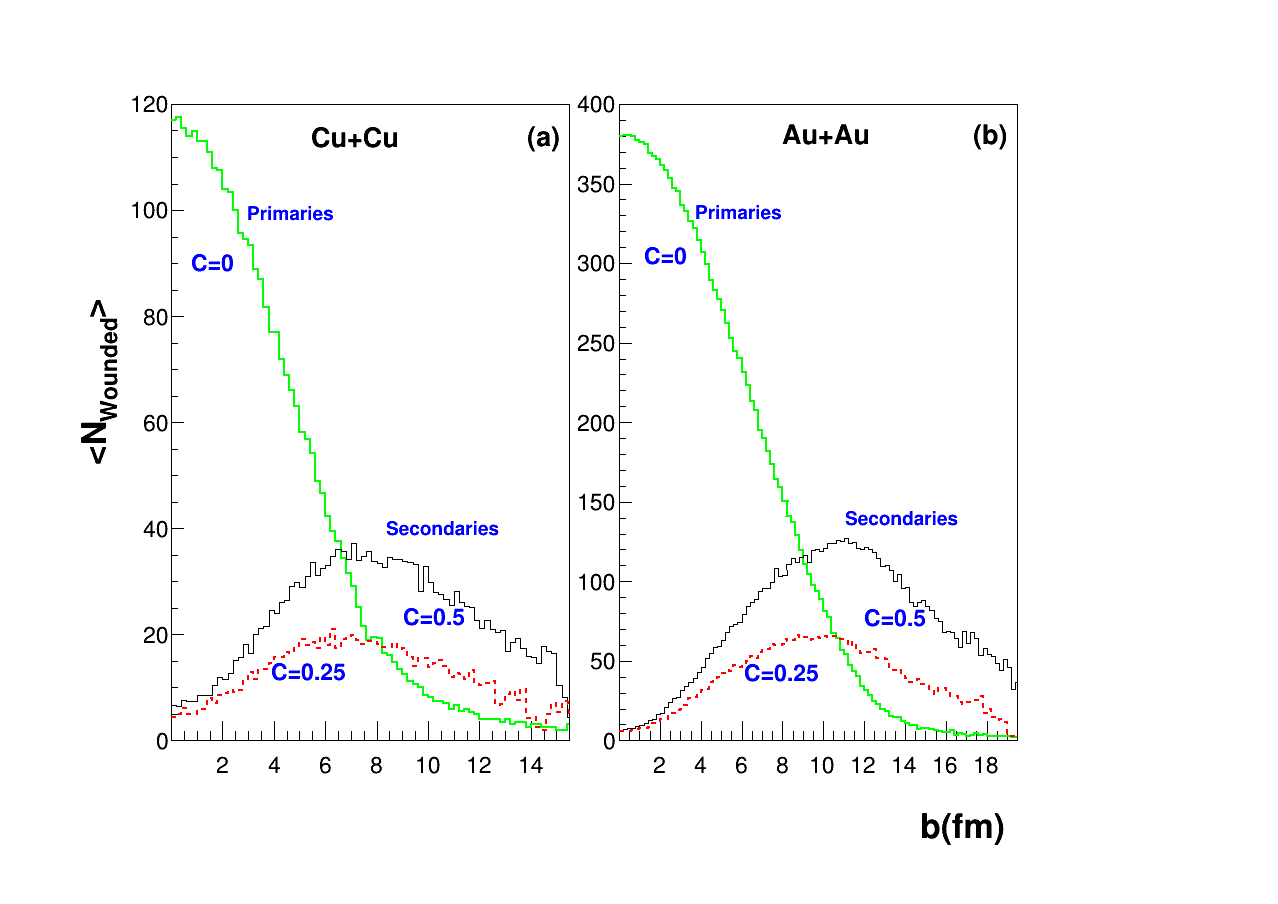}
   		\caption{(COLOR ONLINE)	
   	The average number of interacting nucleons as a function of impact parameter calculated by ImHIJING/Cas code (solid and short-dashed lines) at different cascade strengths ($C$):(a) for $Cu+Cu$ collisions, and (b) for $Au+Au$ collisions at collision energy of $ \sqrt{s_{_{\rm NN}}}=200 $ GeV.		
   		}
   		\label{f3}
   	\end{center}
   \end{figure*}

 Before going any further, it is worthwhile to study the centrality dependence of both primary and accompanying  secondary interacting nucleons for $Cu+Cu$ and $Au+Au$ collisions at $ \sqrt{s_{_{\rm NN}}}=200 $ GeV, see Fig.\ref{f3}. The ImHIJING/Cas calculations show three distinct interaction regions: (i) For impact parameters of $b<2.99$($4.19$)fm, which corresponds to  $0-10\%$ most central  collisions, the ratio of  primary to secondary nucleons is $\sim 11$($\sim 62$) for $Cu+Cu$ ($Au+Au$) collisions; (ii) At $20-30\%$ centrality, which corresponds to $ 4.59<b<5.59$ ($6.39<b<7.99$) for $Cu+Cu$ ($Au+Au$) collisions, the primary and secondary nucleons are found in roughly equal proportions; (iii) In peripheral $Cu+Cu$ ($Au+Au$) collisions at $ 6.19<b<7.59 $ ($8.59<b<10.39$) fm,  which corresponds to $35-50\%$ centrality, the primary to secondary ratio is $\sim 0.5$. Fig.\ref{f3} also indicate that at $40-50$\% centrality, which corresponds to $6.59<b<7.59 (9.39<b<10.39)$ for $Cu+Cu$ ($Au+Au$) collisions, the ImHIJING/Cas calculations with $C=0.25$ cause a significant reduction of secondary interacting nucleons by $\sim 45\%$ relative to the $C=0.5$ case. 
  
The importance of including the collective cascade model at the lowest RHIC energy is demonstrated in Fig.\ref{f4} for central ($6-10\%$) and peripheral ($35-40\%$) $Au+Au$ collisions.  Because jet production is expected to be negligible at this energy, calculations with zero parton shadowing, $s_{q(g)}=0 $, are selected. Solid and short-dashed lines show the ImHIJING/Cas calculations with $C=0.5$ and $C=0$, respectively.  As one can see, the short-dashed lines tend to be higher than the experimental data in the mid-pseudorapidity region at peripheral ($\ge 35-40\%$) collisions. In contrast, the solid lines more closely reproduce experimental data in this region. In other words, adoption of $ \gamma(c) $ and $C=0.5$ in the model calculations are both necessary in order to reproduce the mid-pseudorapidity region. We also note that the cascade effects of the model (solid lines) lead to the broadening of the distributions at large pseudorapidity $ |\eta|>y_{\rm beam} $ as the centrality decreases, in good agreement with the data. 

In Figs.\ref{f5}-\ref{f6} we study the effect of the strength of cascading on $ dN_{ch}/d\eta $ for the reactions under study at the lower RHIC energies. As one can see the strength of cascade effects is shown to be more pronounced in the projectile/target spectator regions, $ |\eta|>y_{\rm beam} $, as expected. In particular, the variant with $C=0.5$ (thick lines) leads to a broadening of the $dN_{ch}/d\eta$ at $y_{\rm beam}<|\eta|<y_{\rm beam}+1$ compared with $C=0.25$ (thin lines) for collision centrality starting from $30-35\%$; see Figs. \ref{f5}h-\ref{f5}j and Figs. \ref{f6}j-\ref{f6}l. We also notice that changing the cascade strength from $C=0.5$ to $C=0.25$ allows us to describe the $ dN_{ch}/d\eta $ in peripheral ($> 30-35\%$)  $Au+Au$  interactions at the lowest RHIC energy. The decreasing strength of cascade going from light to heavy ion collisions implies a mass number dependence. It has also been shown in  Ref.\cite{27} that $C$ depends on both the mass number and beam rapidity.

It should be noted that, the detailed comparisons of the model calculations with the charged particle yield data at $(|\eta|-y_{\rm beam}) >1 $ are not shown in Figs.\ref{f5}h-\ref{f5}j and \ref{f6}j-\ref{f6}l, since both the Fermi motion and nuclear excitation/de-excitation mechanisms are disregarded in ImHIJING/Cas.  
 
From this study of the centrality dependence of $ dN_{ch}/d\eta $ emitted at different collision centralities, for $Au+Au$ and $Cu+Cu$ collisions at the lower RHIC energies, $19.6$ GeV and $22.4$ GeV, respectively, one can conclude that the ImHIJING/Cas has a valid basis for further extrapolations in energy.

At higher collision energy ($\sqrt{s_{_{\rm NN}}}\ge 62.4 $ GeV) the collective cascade model is not expected to fit the measured centrality dependence of $ dN_{ch}/d\eta $ at mid-pseudorapidity, systematically. This is because, in addition to soft interactions, new physics associated with jet production starts to play an important role and is expected to lead to new nuclear dependence of the charged particle production. In particular, charged particles produced in such hard scatterings are created in primary collisions and are centered in a narrow region around mid-pseudorapidity, especially for the $0-10\%$ most central collisions, see Figs. \ref{f7}a-\ref{f7}c. The main two parameters which affect jet  production in high energy heavy ion collisions are jet quenching and parton shadowing. 

The jet quenching parameter, the value of jet energy loss, $ dE/dx $, is determined within HIJING1.383 using the same procedure as proposed in Ref.\cite{kh2};  by studying final state interactions during $Pb+Pb$ collisions  at $\sqrt{s_{_{\rm NN}}}=2.76$ TeV.  We find that  (not shown here) increasing final state interactions, by turning on jet quenching ($ dE/dx>0 $), results in an increase of  the level of $ dN_{ch}/d\eta $ at mid-pseudorapidity; see Fig.\ref{f2} of Ref.\cite{kh2}. Only when the value of  $dE/dx$  is the same as the one for jet production, $p_{0}$,  final state interactions lead to a reproduction of the (double) Gaussian shape of $ dN_{ch}/d\eta $ at mid-pseudorapidity for the reactions under study. 

The value of parton shadowing parameter, $s_{q(g)} $, also affects the $dN_{ch}/d\eta $ yield at mid-pseudorapidity. In high energy heavy ion collisions, depending on nuclear size,  energy and centrality, the low-momentum parton distribution in a nucleon embedded in a nucleus is modified compared to a free nucleon (parton shadowing). Using the studied RHIC data, we find that  $ s_{q(g)} $, which fits the measured centrality dependence of  $dN_{ch}/d\eta $ at mid-pseudorapidity, takes the values of $ s_{q(g)}=0.01 (0.014)$ and $0.07(0.08)$ for $Cu+Cu (Au+Au)$ at $ \sqrt{s_{_{\rm NN}}}=62.4 $ and $200$ GeV, respectively.  

Shown in Figs.\ref{f7} are the plots of ImHIJING/Cas calculations for $Au+Au$ collisions at $\sqrt{s_{_{\rm NN}}}= 62.4 $ GeV using $\alpha_{A} (r_{i} )$ parameterization with $ s_{q(g)}=0.014$ (thick solid lines) and $s_{q(g)}=0$ (thin solid lines). As one can see, parton shadowing lead to  nearly a constant value of charged particle yield at mid-pseudorapidity for peripheral collisions, $ c \ge 35-40 $\%, and then slowly increases as one moves to more central collisions. This clearly indicates that the most significant parameter that affects the charged particle yield at mid-pseudorapidity for $0-10$\% most central $Au+Au(Cu+Cu)$ reactions at  $ \sqrt{s_{_{\rm NN}}} \ge 62.4 $  is $s_{q(g)}$.

It is worthwhile mentioning that the maximum values of $ s_{q(g)}$  for RHIC energies are much smaller than the HIJING2.0 (A MultiPhase Transport, AMPT \cite{8}) estimate of $0.17-0.22$ ($0.1-0.17$) \cite{2, 30}.  However, it is comparable to the maximum value of $ s_{q(g)}=0.1$ that obtained by the experimental data on deep inelastic scattering off nuclear targets \cite{31}. In contrast to HIJING2.0 (AMPT), our estimated $s_{q(g)}$ values do not overlap between RHIC energy ranges, indicating a stronger constraint of $s_{q(g)}$. 

Let us now focus on the role of nucleon shadowing on $ dN_{ch}/d\eta $ for central $Cu+Cu(Au+Au)$ collisions at  $\sqrt{s_{_{\rm NN}}} \ge 62.4$ GeV. This is illustrated in Figs.\ref{f7}a-\ref{f7}c and \ref{f8}, by comparing the model calculations with and without nucleon shadowing along with the measured data. The ImHIJING/Cas calculations are performed using cascade strengths of  $C=0.5$ (solid lines) and $0$ (short-dashed lines). As one can see, there is no difference between the ImHIJING/Cas results, both can describe the mid and smooth fall-off pseudorapidity regions. Such behavior can be related to the difference in energy/momentum: less secondary nucleons are emitted in the $0-3$\% most central collisions with  fractional momenta $ (x_{i(j)}^{\prime \pm}(c)) $ nearly equal to the primary ones (see Figs.\ref{f3}). 

In fact, the largest nucleon shadowing effects are observed in the semiperipheral ($ 20-25 $ \%) and peripheral ($\ge 35-40$\%) interactions, see Figs. \ref{f9} and  Figs.\ref{f10}, which appear when the ratio of secondary to primary nucleons is greater than or equal to one (see Figs. \ref{f3}), for all studied interactions at $ \sqrt{s_{_{\rm NN}}}= 200$ GeV. 

 It is interesting to notice that at $ \sqrt{s_{_{\rm NN}}}=62.4 $ GeV (Figs.\ref{f7}d-\ref{f7}f), there is almost no difference between the calculations with and without nucleon shadowing at the mid-pseudorapidity region for $ c=35-50 $\% . This implies that the threshold of nucleon shadowing effect takes place in $Cu+Cu(Au+Au)$ collisions at $ \sqrt{s_{_{\rm NN}}}=62.4 $ GeV.

\section{\label{sec:level4}SUMMARY AND CONCLUSIONS}
We studied the effects of nucleon shadowing on charged particle  pseudorapidity density  ($ dN_{ch}/d\eta $)  at different collision centrality (from $0-3$\% to $50-55$\%) in $Cu+Cu$ and $Au+Au$ collisions over the  wide energy range of $  \sqrt{s_{_{\rm NN}}}=19.6-200 $ GeV. For this purpose we use HIJING model, with  MSTW2009 parton distribution functions determined from global analysis of hard scattering, and a collective cascade recipe for nucleon shadowing effects. The impacts of these effects on the charged particle spectra for the studied reactions are  investigated and the following conclusions can be drawn:
\begin{enumerate}
		
	\item
	The introduction of an impact parameter dependence of the fractional momentum of wounded nucleons into  the collective cascade recipe leads to the production of $ dN_{ch}/d\eta $ at mid-pseudorapidity for $Au+Au$ collisions at the lowest RHIC energy of  $ \sqrt{s_{_{\rm NN}}}=19.6$ GeV.
	
	\item
	The updated collective cascade model induces both cascade and nucleon shadowing effects.
	 
	\item
	The cascade effects lead to broadening of the high-$ \eta $ tail regions in peripheral ($> 30-35$\%)  $Au+Au(Cu+Cu)$ collisions at $\sqrt{s_{_{\rm NN}}}= $  $19.6$, $22.4$, $62.4$ GeV, in accordance with the measured PHOBOS data.  
	
	\item 
	The nucleon shadowing effects are pronounced only at $\sqrt{s_{_{\rm NN}}}= 200$ GeV, and found to have an impact-parameter dependence to be consistent with $ dN_{ch}/d\eta $ data.
	
	\item
	In contrast to parton shadowing, the nucleon shadowing effects are shown to be lower in central collisions and highest in mid-central and peripheral collisions for the reactions under study at $ \sqrt{s_{_{\rm NN}}} = 200$ GeV.
		 
	\item
	The introduction of nucleon shadowing and its associated cascade effects in HIJING calculations leads to the reproduction of the whole shape of the measured  $ dN_{ch}/d\eta $ for all studied interactions and centralities.
\end{enumerate}

Thus, the present comparisons with PHOBOS data suggest a better agreement with ImHIJING/Cas code. However,  the model calculations are narrower than the data at $\sqrt{s_{_{\rm NN}}}= 200$ GeV. This may imply that a more advanced hadron production model for string fragmentation and/or final state interactions should be implemented in ImHIJING/Cas code.  
This will be further investigated in our future publications.

\section*{\label{sec:level}ACKNOWLEDGMENTS}
The authors would like to thank Prof. V.V. Uzhinskii for checking the ImHIJING/Cas code. Kh. A.-W. would like to thank the members of GEANT4 hadronic group for the hospitality and advice during his visits to CERN.
The authors would like to thank the referees for the comments that improved the quality of the work.

\section*{\label{sec:level}APPENDIX: ImHIJING/Cas's impact parameter cuts}

The centrality cuts are usually derived by relating the average number of participants ($<N_{partic}>$) to the measured $ dN_{ch}/d\eta $ distributions \cite{9}. The unknown $b$-range cuts of ImHIJING/Cas are, however, estimated in this work by mapping the dependence of  $<N_{partic}>$ on impact parameter to the corresponding PHOBOS results of the centrality dependence of $<N_{partic}>$ (see Tables V-VII of \cite{9}). Note that, the concept of $N_{partic}$, the number of nucleons that experiences at least one collision, is the same quantity as the number of primary nucleons introduced in this paper. Tables I and II list the $b$-range values obtained from HIJING Glauber calculations for the studied centrality classes in $Cu+Cu$ and $Au+Au$ collisions at RHIC energies. 
   
\begin{table}[H]
	\caption{The HIJING Glauber calculations of impact parameter cuts corresponding to the centrality bins in $Au+Au$ collision.}
	\vspace{-0.15cm} 
	\begin{center}
		
		\begin{tabular}{c  c  @{\hspace{2em}}c  @{\hspace{2em}}c}
			
			\hline \hline
			\multicolumn{4}{c}{Au+Au} \\  
			\hline
			$ \sqrt{s_{_{\rm NN}}}$ (GeV)  & 200 & 62.4 & 19.6 \\  
			\hline
			centrality &  \multicolumn{3}{c}{$b$-range (fm)} \\
			\cline{2-4}
			0-3 \%   & 0-1.89     & 0-2.09     & 0-1.99 \\ 
			3-6 \%   & 1.89-3.20  & 2.09-3.19  & 1.99-3.09 \\ 
			6-10 \%  & 3.20-4.19  & 3.19-4.29  & 3.09-4.19 \\ 
			10-15 \% & 4.19-5.39  & 4.29-5.29  & 4.19-5.19 \\ 
			15-20 \% & 5.39-6.39  & 5.29-6.19  & 5.19-6.19 \\ 
			20-25 \% & 6.39-7.19  & 6.19-7.09  & 6.19-6.99 \\ 
			25-30 \% & 7.19-7.99  & 7.09-7.69  & 6.99-7.79 \\ 
			30-35 \% & 7.99-8.59  & 7.69-8.59  & 7.79-8.49 \\ 
			35-40 \% & 8.59-9.39  & 8.59-9.19  & 8.49-9.09 \\ 
			40-45 \% & 9.39-9.79  & 9.19-9.89  & 9.09-9.80  \\ 
			45-50 \% & 9.79-10.39 & 9.89-10.19 & 9.80-10.29 \\ 
			\hline \hline
			
		\end{tabular} 
	\end{center} 
\end{table}

\begin{table}[H]
	\caption{The same as Table I, but for $Cu+Cu$ collisions.}
	\vspace{-0.15cm} 
	\begin{center}
		\begin{tabular}{ c c @{\hspace{2em}}c}
			
			\hline \hline
			\multicolumn{3}{c}{Cu+Cu} \\
			\hline
			$ \sqrt{s_{_{\rm NN}}}$ (GeV)  & 200 & 22.4  \\ 
			\hline
			centrality &  \multicolumn{2}{c}{$b$-range (fm)}  \\
			\cline{2-3}

			0-3 \%   & 0-1.99       & 0-1.59    \\ 
			3-6 \%   & 1.99-2.39    & 1.59-2.39    \\ 
			6-10 \%  & 2.39-2.99    & 2.39-2.89    \\ 
			10-15 \% & 2.99-3.59    & 2.89-3.79    \\ 
			15-20 \% & 3.59-4.59    & 3.79-4.39    \\ 
			20-25 \% & 4.59-4.89    & 4.39-4.99    \\ 
			25-30 \% & 4.89-5.59    & 4.99-5.49    \\ 
			30-35 \% & 5.59-6.19    & 5.49-5.99    \\ 
			35-40 \% & 6.19-6.59    & 5.99-6.49    \\ 
			40-45 \% & 6.59-6.99    & 6.49-6.89    \\ 
			45-50 \% & 6.99-7.59    & 6.89-7.39    \\ 
			50-55 \% & 7.59-7.99    & 7.39-7.79    \\
			\hline \hline
			
		\end{tabular} 
	\end{center}          
\end{table}

\begin{center}
	
\end{center} 
 
%---------------figs

\begin{thebibliography}{28}
		\bibitem{1} X. –N. Wang and M. Gyulassy, Phys. Rev. Lett. 68, 1480(1992);
		X. –N. Wang and M. Gyulassy, Phys. Rev. D 44, 3501 (1991).
		
		\bibitem{2} W. –T. Deng, X. –N. Wang, and R. Xu, Phys. Rev. C 83, 014915(2011);
		W. –T. Deng, X. –N. Wang, and R. Xu, Phys. Lett. B 701, 133 (2011); 
		R. Xu, W. –T. Deng and X. –N. Wang, Phys. Rev. C 86, 051901 (2012).
		
		\bibitem{3} K. Werner, F.-M. Liu, and T.Pierog, Phys. Rev. C 74, 044902(2006).
		
		
		\bibitem{4} S. A. Bass et al., Prog. Part. Nucl. Phys. 41, 255(1998).
		
		\bibitem{5} M. Mitrovski, T. Schuster, G. Graf, H. Petersen, and M. Bleicher, Phys. Rev. C 79, 044901 (2009).
		
		\bibitem{6} T. Hirano, U.W. Heinz, D. Kharzeev, R. Lacey and Y. Nara, Phys. Rev. Lett. B 636, 299 (2006).
		
		\bibitem{7} A. Adil, H. –J. Drescher, A. Dumitru, A. Hayashigaki and Y. Nara, Phys. Rev. C 74, 044905 (2006);  H. –J. Drescher and Y. Nara, Phys. Rev. C 76, 041903(2007).
		
		\bibitem{8} Z. W. Lin, C. M. Ko, B. –A. Li, B. Zhang, S. Pal, Phys. Rev. C 72, 064901 (2005).
		
		\bibitem{dis1} M. Arneodo, Phys. Rep. 240, 301 (1994).
		 
		\bibitem{dis2} P. R. Norton, Rep. Prog. Phys. 66, 1253 (2003).
		
		\bibitem{pdf1} K. J. Eskola, H. Paukkunen and C. A. Salgado, J. High Energy Phys. JHEP0904 (2009)065.
		
		\bibitem{pdf2} M. Hirari, S. Kumano and T-H Nagai, Phys. Rev. C 76, 065207 (2007).
		
		\bibitem{pdf3} D. de Florian, R. Sassot, P. Zurita and M. Stratmann, Phys. Rev. D 85, 074028 (2012).
		
		\bibitem{pdf4} S. A. Kulagin, J. of Phys.: Conference Series 762, 012072 (2016).
		
		\bibitem{kh1} K. Abdel-Waged and N. Felemban, Phys. Rev. C 91, 034908 (2015).

		\bibitem{kh2} K. Abdel-Waged and N. Felemban, Phys. Rev. C 93, 024910 (2016).
		
		\bibitem{martin} A. D. Martin, W. J. Stirling, R. S. Thorne, and G. Watt, Eur. Phys. J. C 63, 189 (2009).
		
        \bibitem{9} B. Alver et al., (PHOBOS collaboration), Phys. Rev. C 83, 024913 (2011).
        
	    \bibitem{kh3} Kh. Abdel-Waged and V.V. Uzhinskii, Phys. Atom. Nucl. 60, 828 (1997).
	    
	    \bibitem{kh4} Kh. Abdel-Waged and V.V. Uzhinskii, J. Phys. G: Nucl. Phys. 24, 1723(1998).
		
		\bibitem{26} Kh. Abdel-Waged, Nuha Felemban and V.V. Uzhinskii, Phys. Rev. C 84, 014905 (2011).
		
		\bibitem{27} A. S. Galoyan, A. Ribon, and V. V. Uzhinsky, JETP letters 102, 6, 324 (2015). 
		
		\bibitem{28} W. Broniowski and W. Florkowski, Phys. Rev. C 65, 024905 (2002).
		
		
		\bibitem{30} Subrata Pal and Marchus Bleicher, Phys. Lett. B 709, 82-86 (2012).
		
		\bibitem{31} S. –Y. Li, X. –N. Wang, Phys. Lett. B527, 85  (2002).
		
		\bibitem{32} T. Pierog, I. Karpenko, J. M. Katzy, E. Yatesenko, K. Werner, Phys. Rev. C 92 (3) 034906 (2015).
		
	\end{thebibliography}
\end{document}